\documentclass[iop,apj,tighten]{emulateapj}

\usepackage[english]{babel}
\usepackage[utf8]{inputenc}
\usepackage{amsmath}
\usepackage{graphicx}
\usepackage[colorinlistoftodos]{todonotes}
\usepackage{natbib} 

\begin{document}
\title{Analytic Asymptotic Solution to Spherical Relativistic Shock Breakout}
\author{Almog Yalinewich\altaffilmark{1} and Re'em Sari\altaffilmark{1}}
\affil{$^1$Racah Institute of Physics, the Hebrew University, 91904, Jerusalem, Israel}

\date{\today}

\begin{abstract}
We investigate the relativistic breakout of a shock wave from the surface of a star. In this process, each fluid shell is endowed with some kinetic and thermal energy by the shock, and then continues to accelerate adiabatically by converting thermal energy into kinetic energy. This problem has been previously studied for a mildly relativistic breakout, where the acceleration ends close to the surface of the star. The current work focuses on the case where the acceleration ends at distances much greater than the radius of the star. We derive an analytic description for the hydrodynamic evolution of the ejecta in this regime, and validate it using a numerical simulation. We also provide predictions for the expected light curves and spectra from such an explosion. The relevance to astrophysical explosions is discussed, and it is shown that such events require more energy than is currently believed to result from astrophysical explosions.
\end{abstract}

\section{Introduction}
A supernova deposits a huge amount of energy in the bowels of a star. From that hot spot emerges a radiation dominated shock wave which propagates outward. Close to the stellar edge, i.e. the stellar atmosphere, the density decreases sharply. When the shock wave approaches the stellar surface, it accelerates, perhaps even to relativistic velocities. At a later point, when the shock wave is even closer to the stellar edge, radiation begins to leak out. This is called shock breakout, and it is the first electromagnetic signal from a supernova explosion. 

The hydro - radiation evolution of shock breakout has been studied extensively, both analytically and numerically. The evolution can be divided into two phases. First comes the planar phase, where the displacement of a fluid element is much smaller than its radius (i.e. distance from the core). Later comes the spherical phase, where the displacement is much larger than the initial radius. In the Newtonian regime, analytic studies probed both the planar phase \citep{piro_et_al_2010} and the spherical phase \citep{nakar_sari_newtonian_breakout_2010}, while in the ultra relativistic regime only the planar phase was studied \citep{nakar_sari_relativistic_breakout_2012}. 

Analysis of the breakout relies on a relation between the Lorentz factor of a certain fluid element attains immediately after it is shocked $\gamma_s$, and its terminal Lorentz factor $\gamma_t$. By terminal we mean after the volume of a fluid element doubled its size many times, and exhausted its thermal energy. In the case of a planar shock, it was shown that this relation is $\gamma_t \approx \gamma_s^{1+1/\sqrt{3}}$ \citep{johnson_mckee_1971}. One way of deriving this relation is by using the conservation of the relativistic forward Riemann invariant. The same Riemann invariant, however, is not conserved when the flow is spherical. Luckily, it is possible to obtain a modified version of the Riemann invariant that is conserved behind an outward moving shock in the spherical regime as well \citep{oren_sari_relativistic_2009}. In this paper we use this modified Riemann invariant to obtain a relation between the initial (shocked) and terminal Lorentz factors for a spherical breakout. The analysis presented here is valid for fluid elements which only attain their terminal values when the displacement was much larger than the radius of the star. These are fluid elements which originally reside very close to the edge of the star.

\section{Overview of Shock Breakout Theory} \label{sec:overview}
In this section we summarize the theory of shock breakout discussed in previous works
\citep{sakurai_1960,johnson_mckee_1971,katz_et_al_2010,nakar_sari_newtonian_breakout_2010,nakar_sari_relativistic_breakout_2012}.
We assume that a spherical progenitor with mass $M$ and radius $R$ explodes with energy $E$. We assume that the explosion starts from a hot spot at the very center of the progenitor. 
The shock wave travels outward, and when it reaches the stellar atmosphere it accelerates due to the declining density profile.
We assume that very close to the edge of the progenitor the number density scales as $n_a \propto x^{\omega}$, where $x = R - r$ and $r$ is the radial coordinate (distance from the center of the progenitor). 
In order for the total mass to be of order $M$, and assuming the star is made up of hydrogen the number density has the form
\begin{equation}
n_a \approx \frac{M}{m_p R^3} \left(\frac{x}{R} \right)^{\omega}
\end{equation}
where $m_p$ is the mass of the proton. Initially, the shock is not relativistic, and the velocity increases like $v \propto n_a^{-\mu'}$ \citep{sakurai_1960}. The velocity profile is given by \citep{matzner_mckee_1999}
\begin{equation}
v \approx \sqrt{\frac{E}{M}} \left( \frac{x}{R} \right)^{-\mu' \omega}.
\end{equation}
The shock may break out in this, non relativistic phase, but in this work we are interested in cases where the shock keeps accelerating to relativistic velocities. In this stage the Lorentz factor scales with the density as $\Gamma \propto n_a^{-\mu}$. The values of both power law indices are very close: $\mu \simeq 0.23$ \citep{sari_self_simialr_relativistic_shock_2005} and $\mu' \simeq 0.19$ \citep{sakurai_1960}. For simplicity we assume $\mu = \mu'$. The relativistic shock trajectory can be smoothly connected to the non - relativistic trajectory by the condition that $\Gamma \approx 1$ when $v \approx c$, hence
\begin{equation}
\Gamma \approx \sqrt{\frac{E}{M c^2}}\left( \frac{x}{R}\right)^{-\mu \omega}.
\end{equation}
In the absence of pair production, shock acceleration stops at optical depth of order unity, because from that point on photons can leak out. Assuming the dominant scattering process is Compton scattering, the depth of the breakout shell is given by
\begin{equation}
x_0 = R \left( \frac{M}{m_p} \frac{\sigma}{R^2}\right)^{-\frac{1}{\omega +1}}
\end{equation}
The condition for the breakout shell to be relativistic is
\begin{equation}
\sqrt{\frac{E}{M c^2}} \left( \frac{M}{m_p} \frac{\sigma}{R^2} \right)^{\frac{\mu \omega}{\omega +1}} > 1.
\end{equation}
For $\omega = 3$, this condition can be written as
\begin{equation}
\frac{E}{10^{51}\, \rm{erg}} > 0.8 \left( \frac{M}{M_{\odot}}\right)^{0.7} \left( \frac{R}{R_{\odot}}\right)^{0.6}
\end{equation}
Fluid shells continue to accelerate even after the shock reaches the edge of the progenitor and disappears. If a fluid element finishes accelerating before doubling its radius, then the relation between the shocked Lorentz factor $\gamma_s$ and the terminal Lorentz factor $\gamma_t$ is \citep{johnson_mckee_1971}
\begin{equation}
\gamma_t \approx \gamma_s^{1+\sqrt{3}}. \label{eq:planar_breakout}
\end{equation}
This is referred to as planar breakout.
Right after a fluid element is shocked, the Lorentz factor evolves as \citep{pan_sari_2006}
\begin{equation}
\gamma \approx \gamma_s \left( \frac{t}{x/c} \right)^{\frac{\sqrt{3}-1}{2}}
\end{equation}
where $t$ is the lab frame (not comoving) time measured from breakout. The time it takes a fluid element to reach the terminal planar Lorentz factor is therefore
\begin{equation}
t_t \approx \frac{x}{c} \gamma_s^{3+\sqrt{3}}.
\end{equation}
The time it takes a shell to double its radius is
\begin{equation}
t_d \approx \frac{R}{c}
\end{equation}
The approximation of a planar breakout is valid if $t_d>t_t$. The breakout shell is the fastest shell, so if the planar approximation is valid for the breakout shell, it remains valid for inner shells. The condition for that is
\begin{equation}
1 > \left(\frac{E}{M c^2} \right)^{\frac{3+\sqrt{3}}{2}} \left( \frac{M}{m_p} \frac{\sigma}{R^2} \right)^{1-\mu \omega \left(3 + \sqrt{3} \right)}.
\end{equation}
If $\omega = 3$, then the inequality above can be written as
\begin{equation}
\frac{E}{10^{51} \, \rm{erg}} < 12 \left( \frac{M}{M_{\odot}}\right)^{0.8} \left( \frac{R}{R_{\odot}}\right)^{0.4} \label{eq:sbo_condition}
\end{equation}
The comoving temperature of the shock is about $T_i' \approx 200 \, \rm{keV}$, regardless of the shock Lorentz factor, due to the production of pairs \citep{katz_et_al_2010}. These pairs also blow up the opacity, so radiation can only escape when the temperature drops and the pairs disappear. This happens at a temperature of about $T_{th}' \approx 50 \, \rm{keV}$.
The energy of the breakout shell when it is shocked is
\begin{equation}
E_{bs} \approx m_p n x R^2 c^2 \gamma_s^2 \approx E \left( \frac{M}{m_p} \frac{\sigma}{R^2}\right)^{-1+\frac{2 \mu \omega}{\omega +1}}
\end{equation}
At that moment the temperature of the shell is $T_i'$ and its Lorentz factor is $\gamma_s$. When it stops accelerating and becomes transparent, its temperature is $T_{th}'$ and its Lorentz factor is $\gamma_t$, so its energy at this point is
\begin{equation}
E_{bt} \approx E_{bs}\frac{T_{th}'}{T_i'} \frac{\gamma_t}{\gamma_s} \approx E \left( \frac{E}{M c^2} \right)^{\sqrt{3}/2} \left( \frac{M}{m_p} \frac{\sigma}{R^2} \right)^{-1 + \frac{\left( 2 - \sqrt{3}\right) \mu \omega}{\omega + 1}}
\end{equation}
This energy is released almost instantaneously in the lab frame, but when it arrives to an observer it is stretched over a time period of $t_{ob} \approx \frac{R}{\gamma_t^2 c}$. The observed temperature of the emitted light is $T_{ob} \approx T_{th}' \gamma_t$. Eliminating the Lorentz factors yields a closure relation between the energy, temperature and duration of the burst
\begin{equation}
t_{ob} \approx 20 \, {\rm{s}} \left( \frac{E}{10^{46} \, \rm{erg}}\right)^{1/2} \left(\frac{T}{T_{th}'} \right)^{-\frac{9+\sqrt{3}}{4}}
\end{equation}
Because of pair production, the shock increases the opacity of fluid shell it sweeps. This mean that some radiation may originate in shells outside the breakout shell, where the apparent temperature is higher. The energy of a ``superior" shell increases with its thickness $E_s \approx n_a x R^2 m_p c^2 \gamma_s^2 \frac{\gamma_t}{\gamma_s} \frac{T_{th}'}{T_i'}$, and the time interval over which it arrives to an observer is $t \approx \frac{R}{\gamma_t^2 c}$. The apparent luminosity (in the observer frame) is therefore
\begin{equation}
L \approx \frac{E c}{R} \left( \frac{E}{M c^2}\right)^{\frac{-2 \mu  \omega +\omega +1}{2 \mu  \omega }} \left(\frac{c t}{R} \right)^{\frac{-3 \sqrt{3} \mu  \omega -4 \mu  \omega +\omega +1}{2 \sqrt{3} \mu  \omega
   +2 \mu  \omega }}
\end{equation}
For $\omega = 3$ we can rewrite the equation above as
\begin{equation}
L \approx 2.2 \cdot 10^{44} \, {\rm{\frac{erg}{s}}} \left( \frac{E}{10^{51} \, \rm{erg}}\right)^{2.9} \left( \frac{M}{M_{\odot}}\right)^{-1.9} \times
\end{equation}
\begin{equation*}
\times \left( \frac{R}{R_{\odot}}\right)^{-0.38} \left( \frac{t}{1\, \rm{s}}\right)^{-0.62}
\end{equation*}
Each shell has a different temperature, and we assume that the photons it emits are mostly thermal $h \nu \approx k T$. Since $t \propto \gamma^{-2} \propto T^{-2}$, the integrated spectrum is given by
\begin{equation}
\nu F_{\nu} \propto t L \propto t^{0.38} \propto \nu^{-0.76}
\end{equation}
where the numerical values are for $\omega = 3$.

\section{Spherical Relativistic Breakout}
We now deal with the case where the breakout shell travels to a radius much larger than the radius of the star before the acceleration stops. In this case equation \ref{eq:planar_breakout} no longer holds. Each fluid shell is characterized by four variables: density $\rho$, pressure $p$, Lorentz factor $\gamma$ and radius $r$.
We want to relate the properties of a fluid shell right after it was shocked to the properties of the same shell when the acceleration ends. We denote by subscript $s$ variables related to the shocked state, and subscript $t$ to the terminal state. We would like to express the hydrodynamic variables in terms of the shocked Lorentz factor. The Taub equations relate the shocked Lorentz factor to the downstream density and pressure
\begin{equation}
n_s \approx n_a \gamma_s \label{eq:first}
\end{equation}
and
\begin{equation}
p_s \approx m c^2 n_a \gamma_s^2
\end{equation}
where $m$ is the rest mass of the particles and $c$ is the speed of light. The initial radius of the shell is simply the radius of the star
\begin{equation}
r_s \approx R.
\end{equation}
Conservation of entropy yields
\begin{equation}
\frac{p_s}{n_s^{4/3}} = \frac{p_t}{n_t^{4/3}}.
\end{equation}
The acceleration ends when the thermal pressure is comparable to the rest mass energy
\begin{equation}
p_t \approx n_t m c^2.
\end{equation}
In the planar case $r_t \approx r_s \approx R$ and this set of equations is closed using the conservation of the relativistic Riemann invariant \citep{johnson_mckee_1971}
\begin{equation}
J\left(p_s, \gamma_s \right) = J \left(p_t, \gamma_t \right)
\end{equation}
where
\begin{equation}
J\left(p,\gamma \right) = \ln \gamma + \frac{\sqrt{3}}{4} \ln p.
\end{equation}
However, the Riemann invariant is only conserved in planar, and not spherical flow. \cite{oren_sari_relativistic_2009} obtained a modified Riemann invariant that is conserved in a spherical flow
\begin{equation}
J' \left(p, \gamma, r \right) = J\left(p, \gamma \right) + \left(\sqrt{3}-1 \right) \ln r
\end{equation}
The final relation can be obtained by keeping track of how much a fluid element expands. For this part we assume the following ansatz for the terminal Lorentz factor
\begin{equation}
\gamma_t = C_{\gamma} \gamma_s^{\sigma} \left( \frac{E}{M c^2}\right)^{\psi} \label{eq:gamma_t_rel}.
\end{equation}
If the initial width of some fluid element is $ dx $, the terminal width would be 
\begin{equation}
dx_t \approx r_t \cdot dx \frac{d\beta_t}{dx} \approx \cdot dx \cdot \gamma_s^{-2 \sigma -\frac{1}{\mu \omega}} \left( \frac{E}{Mc^2}\right)^{2\psi}.
\end{equation}
The terminal density scales as 
\begin{equation}
n_t \approx n_0 \frac{R^2 dx}{\gamma_t r_t^2 dx_t} \propto r_t^{-3} \gamma_s^{\sigma - \frac{1}{\mu} + \frac{1}{\mu \omega}} \left(\frac{E}{M c^2} \right)^{\psi}. \label{eq:density_rel}
\end{equation}
Putting together equations \ref{eq:first} - \ref{eq:density_rel},  we get

\begin{equation}
\sigma = \frac{2 \left(\sqrt{3}-2\right) \mu  \omega +\frac{4}{3}  (5
   \mu  \omega +2)-2}{\left(20/3 +\sqrt{3} -5\right)
   \mu  \omega },
\label{eq:sigma_expr} \end{equation}

\begin{equation}
\psi = -\frac{1}{5 \mu  \omega +3 \sqrt{3} \mu  \omega } \label{eq:psi_expr}
\end{equation}
and $C_{\gamma}$ is a dimensionless constant of order unity which can be calibrated from simulations (see next section). For $\omega = \frac{3}{2} $ and $\eta = \frac{4}{3} $ we get $\sigma \simeq 2.36$ and $\psi \simeq -0.28$, and for $\omega=3$ and $\eta=\frac{4}{3}$ we get $\sigma \simeq 2.1$ and $\psi \simeq -0.14$. We can also get the radius at which the acceleration stops

\begin{equation}
\frac{r}{R_*} \propto \gamma_s ^{\lambda} \left( \frac{E}{M c^2} \right)^{\varpi}
\end{equation}
where
\begin{equation}
\lambda = \frac{\left(4\sqrt{3}  + 6 \right) \omega  -  10 \left(1 + \sqrt{3}\right)}{\omega  \left(5 + 3\sqrt{3}\right)}
\end{equation}
and 
\begin{equation}
\varpi = \frac{1 + 1/\sqrt{3}}{3 \mu \left( 3 + 5\right)}
\end{equation}
For $\omega = \frac{3}{2}$ and $\eta = \frac{4}{3}$ we get $\lambda \simeq 0.49$ and $\varpi \simeq 0.39$, and
for $\omega = 3$ and $\eta = \frac{4}{3}$ we get $\lambda \simeq 0.88$ and $\varpi \simeq 0.19$.

\section{Simulation}
\begin{figure}[ht!]
\begin{center}
\includegraphics[width=1.0\columnwidth]{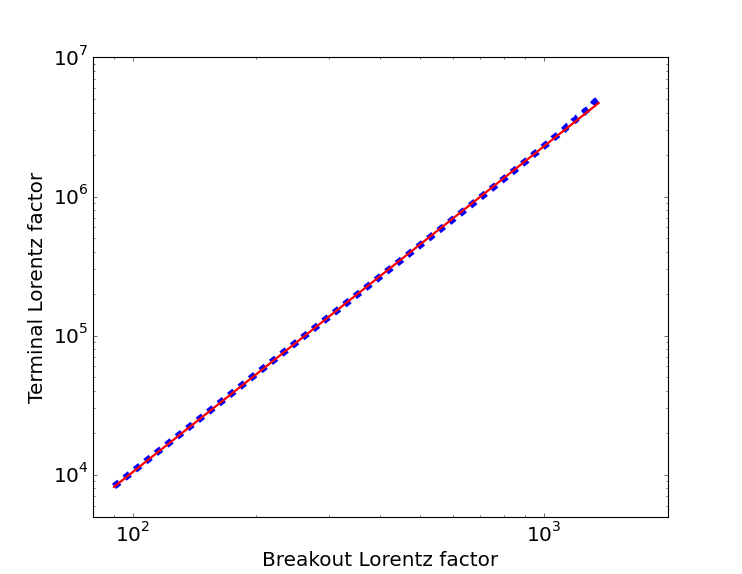}
\caption{Lorentz factors of computational cells at breakout versus the Lorentz factors of the same cells at a much later time. Blue represents data obtained from the simulation, and red is a power law fit. The power law index inferred from the simulation is 2.2, while the analytic value is 2.1.} The details of the simulation are in the text. \label{fig:rel_sph_bo}
\end{center}
\end{figure}

In order to verify our results we ran a one dimensional, lagrangian, relativistic simulation based on \citet{sari_kobayashi_piran_1999} and \citet{2000A&A...358.1157D}. Our computational domain is $10^{-3} < r < 1-10^{-4}$. The stellar edge is at $r_s = 1$. We use a logarithmic grid, such that for each cell, the ratio between its width and the distance to the edge is constant and equal to 0.005. The initial density profile is a power law of the distance to the edge $\rho = \left( r_s - r\right)^{\omega} $, where $\omega = 3$. The initial pressure is almost zero throughout the domain, except for a small ``hot spot" where it is finite

\begin{equation}
p = \left\{
\begin{array}{ll}
      5 \cdot 10^{7} & 0.01 > r \\
      10^{-30} & r \ge 0.01 \\
\end{array} 
\right.
\end{equation}
We let the simulation run to time $t = 10^{6}$. We took two snapshots: one at breakout time, and another at the final time. In figure \ref{fig:rel_sph_bo} we plot the terminal Lorentz factor as a function of breakout Lorentz factor for each computational cell. By fitting a power law to the this plot we obtain a power law index of 2.2, which is close to the theoretical prediction 2.1 (relative difference of about 5\%).

These results can be used also to calibrate the dimensionless coefficient $C_{\gamma}$ from equation \ref{eq:gamma_t_rel}. In this simulation $\frac{E}{M c^2} \approx 10^3$, so $C_{\gamma} \left(\omega = 3, \eta=\frac{4}{3} \right) \approx 0.9$.

\section{Light Curves}
In this section we follow the derivation in \citep{nakar_sari_relativistic_breakout_2012}, but with a slight change. Instead of $\gamma_t \approx \gamma_s^{\sqrt{3}+1}$, the relation between the initial and final Lorentz factors is $\gamma_t \approx \left( \frac{E}{M c^2} \right)^{\psi} \gamma_s^{\sigma}$, with $\sigma$ from equation \ref{eq:sigma_expr} and $\psi$ from equation \ref{eq:psi_expr}. The energy of each shell, when it is shocked, scales as

\begin{equation}
E_i \propto \gamma_s^{-\frac{1+\omega \left(1-2 \mu \right)}{\omega+1}}
\end{equation}
where $\gamma_s$ is the Lorentz factor of the shocked material. The emitted energy when the fluid element becomes transparent is $E_f = E_i \frac{\gamma_t}{\gamma_s}$. This energy is emitted within a time interval

\begin{equation}
t_{obs} \approx \frac{r_t}{c \gamma_t^2}.
\end{equation}
The dependence of the bolometric luminosity on time is therefore given by
\begin{equation}
\frac{d\ln L}{d \ln t} = \frac{1}{2 \mu \omega \sigma} \left(- 3 \mu \omega \sigma - \mu \omega + \omega + 1\right)
\end{equation}
which, for $\omega = 3$, comes out to be $\frac{d \ln L}{d \ln t} = -0.36$ instead of -0.62 in the planar case.
The rest frame temperature of each shell is constant $T_{th}' \approx 50\,\rm{keV}$ \citep{katz_et_al_2010}. The temperature in the observer frame is $T_{th}' \cdot \gamma_t$, hence
\begin{equation}
\frac{d \ln T}{d \ln t} = -\frac{1}{2 - \lambda / \sigma}
\end{equation}
The spectrum is obtained using the same rationale as in section \ref{sec:overview}
\begin{equation}
\nu F_\nu \propto t L \propto t^{0.64} \propto \nu^{-0.8}
\end{equation}
whereas the corresponding value for a planar breakout is -0.74.

It is also possible to obtain closure relations, in the same way as was done in \citep{nakar_sari_relativistic_breakout_2012}. For typical values $\omega = 3$ and $\eta = 4/3$ we get

\begin{equation}
t_{obs} \approx {7 \, \rm{s}} \left( \frac{E_{bo}}{10^{46} erg} \right)^{0.59} \left( \frac{T_{bo}}{T_{th}'} \right)^{-2.21} \left( \frac{M}{M_{\odot}}\right)^{-0.09}. \label{eq:spherical_closure}
\end{equation}
Since the minimum terminal Lorentz factor at which spherical breakout occurs is \citep{nakar_sari_relativistic_breakout_2012}
\begin{equation}
\gamma_{f,s} \approx 30 \left(\frac{M}{5 M_{\odot}} \right)^{0.14} \left( \frac{R}{5 R_{\odot}}\right)^{-0.27},
\end{equation} 
the minimum temperature is $1.5 \, \rm{MeV}$. We recall that for a planar breakout \citep{nakar_sari_relativistic_breakout_2012}

\begin{equation}
t_{obs} \approx {20 \, \rm{s}} \left( \frac{E_{bo}}{10^{46} \, \rm{erg} }\right)^{1/2} \left(\frac{T_{bo}}{T_{th}'} \right)^{-\frac{9+\sqrt{3}}{4}}. \label{eq:planar_closure}
\end{equation}
The exponents in the spherical closure relation (equation \ref{eq:spherical_closure}) are very similar to that of the planar closure relations (equation \ref{eq:planar_closure}). Therefore the planar closure relations can be used as a good approximation even in the spherical case.

\section{Discussion}
We examined shock breakout from a star in the regime where the ejecta is relativistic, and exhausts its thermal energy at a radius much larger than that of the star. We found a relation between the Lorentz factor of each shell at breakout and the Lorentz factor after the exhaustion of thermal energy. We verified our result using a numerical simulation. Finally, we used these results to predict the light - curves and spectrum from such an explosion.

Spherical breakout occurs only for sufficiently high energies (equation \ref{eq:sbo_condition}). 
Typical supernova energies are around $10^{51}\,\rm{erg}$, and typical masses of supernova progenitors are of the order of $M_{\odot}$. Substituting these values into equation \ref{eq:sbo_condition} yields that the radius must be smaller than a few hundred kilometers. The only astrophysical objects that satisfy these conditions are neutron stars and black holes, and only the former can explode. However, such strong explosions in neutron stars will produce neutrinos. Neutrinos have a much smaller cross section than photons, so they will escape at a much earlier stage, before the shock manages to reach high Lorentz factors. This problem is further aggravated by the degeneracy, which increases the effective mean free path for scattering. 

\bibliographystyle{yahapj}
\bibliography{almogyalinewich.bib}

\end{document}